# Trends in Russian research output indexed in Scopus and Web of Science

Henk F. Moed (*), Valentina Markusova[1] (**) and Mark Akoev (***)

* Sapienza University of Rome, Rome, Italy. Email: hf.moed@gmail.com

** All Russian Institute for Scientific &Technical Information of the Russian Academy of Sciences (VINITI), Moscow, Russia. Email: markusova@viniti.ru.

*** Ural Federal University, Scimetrics Lab, Ekaterinburg, Russia. Email: m.a.akoev@urfu.ru

**Abstract**

Trends are analysed in the annual number of documents published by Russian institutions and indexed in Scopus and Web of Science, giving special attention to the time period starting in the year 2013 in which the *Project 5-100* was launched by the Russian Government. Numbers are broken down by document type, publication language, type of source, research discipline, country and source. It is concluded that Russian publication counts strongly depend upon the database used, and upon changes in database coverage, and that one should be cautious when using indicators derived from WoS, and especially from Scopus, as tools in the measurement of research performance and international orientation of the Russian science system.

**Keywords**

Scopus, WoS, article, review, proceedings, compound annual growth rate, citation, Russia, research discipline, Project 5-100, publication language

## 1. Introduction

During the past two decades, major changes took place in the research system of the Russian Federation. A series of government initiatives was launched, aimed to increase competition among Russian universities, and enhance their international status and visibility. One of the most important initiatives is the *Project 5-100* to improve the prestige of Russian higher education and bring at least five universities from among the project participants into the hundred best universities in the world according to three world university rankings, namely the Shanghai Ranking (Academic Ranking of World Universities), Times Higher Education (THE) Ranking, or QS (Quacquarelli Symonds) ranking.

In the assessment of the success of these initiatives, especially of the *Project 5-100*, science indicators are important assessment tools. One important subclass consists of bibliometric indicators of research output and impact, derived from large, multidisciplinary, bibliographical databases Web of Science and Scopus (Van Raan, 2004). In the compilation of the three above mentioned university ranking systems such indicators play a key role.

[1] This paper is partly supported by the Russian Foundation of Basic Research (Grant: 17-02-00157).

Insight into differences in source coverage policies of bibliographical databases and their effects upon bibliometric indicators is essential for a proper interpretation and use of bibliometric indicators in research assessment. This paper aims to contribute to such insight, by presenting a comparative, longitudinal study of the coverage of the publication output from Russian institutions in Scopus and the Web of Science (WoS).

A striking key observation that served as a starting point for the analyses presented in this paper is that the number of documents published from Russian institutions and indexed in Scopus strongly increased during 2000-2016, and that it shows during the past four years even an exponential growth. This result was presented by the first author of the current paper at the NEICON conference organised in 2017 in Jesolo, Italy (NEICON, 2017). The following questions were raised, which are denoted as the core questions in this paper:

i. How can the large increase in the number of documents from Russia indexed in Scopus be explained? Which factors are responsible?
ii. How does Scopus compare with Web of Science (WoS) in this respect?
iii. What is the annual trend in the number of documents from Russian institutions in Scopus and WoS, disaggregated by document type, publication language, and research discipline?
iv. How does the trend in number of documents from Russia indexed in Scopus and WoS compare to those calculated for documents from fellow BRIC countries Brazil, China and India, and to the ten countries with the largest publication output in 2016?
v. To what extent can the observed patterns be attributed to an increase in research performance, and to what extent to changes in the source coverage of the database?

The current paper is of a methodological nature. It does *not* provide a comprehensive bibliometric analysis of Russian research performance. Being interested primarily in the effect of changes or differences in source coverage of bibliographic databases upon the measurement of Russian publication output, it focuses on publication counts, and does not systematically assess citation impact, an aspect of great importance in fully-fledged bibliometric assessment studies.

The paper does *not* present analyses at the level of individual institutions. From the point of view of the research questions addressed above it is not necessary to do so. The paper aims at providing relevant background knowledge about the bibliometric measuring devices Scopus and WoS, and in this way *facilitate* a proper assessment of Russian institutions, rather than conducting such an assessment itself. A necessary condition for any assessment is that it is based on accurate, verified information on the units of assessment. The current authors do not have such high quality information on Russian institutions.

The structure of this paper is as follows. Section 2 presents an overview of the major trends in the Russian science system during the past two decades. Section 3 discusses the data collection carried out in the study. The paper analyses in Section 4 in separate sub-sections the annual trends in breakdowns of Scopus and WoS publication counts by document type, publication language, type of publication, research discipline and country. The last sub-section of Section 4 focuses on the percentage of documents from Russian institutions relative to the global publication output indexed in the two databases, and examines whether this percentage reached in 2016 the value 2.44, one of the central goals of Russian science policy formulated in 2012, and further discussed in Section 2. Finally, Section 5 discusses the implications of the outcomes of the bibliometric analyses presented in Section 4 for answering or illuminating the paper's core question.

## 2. Background

The Russian research system is very different from that of other developed nations as was described by Graham (1995), Wilson (2004), and Karaulova (2016). During the last 20 years the Russian science community has been struggling to hold a leading place on the international science stage. Reform of two main Russian academic sectors, namely the Russian Academy of Sciences (RAS) and the Higher Education Sector (HES), has been going on for the last thirteen years, with the government shifting its attention and financial resources toward the HES. According to the Russian Ministry of Education and Science, HES has in 2017 under its auspice 769 universities and 692 branches in various cities (MICCEDU, n.d.). Among these only 506 conduct basic research (Mindeli, 2013). In 2006, the government ordered a reduction by 20% of the research personnel of the Russian Academy of Sciences (there are about 400 organizations under the auspices of the RAS), the leading research entity in the country. Simultaneously, the government released a decree to set up a Federal University (FU) in each of the ten Federal districts. In 2008, the title "National Research University" (NRU) was awarded to 28 universities after a two-tier competition. The goal of an NRU is to focus on the transfer of knowledge to industry. A few publications in "Nature" discussed a shift toward fostering research in the higher education sector (Schiermeier, 2007; 2010; 2012).

In May 2012, President of the Russian Federation V. Putin released decree № 599, in which he set the goal: "that the Russian share of research output (RO) has to reach 2.44% of the global RO, and five Russian universities have to be among the top hundred universities included in one of three world ranking systems in 2015" (Decree 599, 2012). A new project, denoted as *Project 5-100* started in January 2014 when funding for 2013 was transferred to a selected group of universities.

In June 2013, a new bill introduced and adapted quickly by Russian Parliament, related to the drastic reform of three government academies: the Russian Academy of Sciences (RAS), the Russian Academy of Agriculture and the Russian Academy of Medical Sciences. This reform caused controversy and strong resistance by the Russian research community. On November 1, 2013 D. Medvedev - the Prime Minister of the Russian Federation - released the decree № 979 ordering to include in any research organization's evaluation the following bibliometric indicators: number of papers, citation score and impact factor by Web of Knowledge or Scopus. (see for instance http://www.ras.ru/news/shownews.aspx?id=613a30f8-1475-4d9a-a6a3-75df1501be7a)

The bibliometric performance of the RAS and the HES played a very important role in this reform (Ivanov et al., 2014). Recent papers published in *Scientometrics* (Karaulova et al. 2016) and in the *Herald of Russian Academy of Sciences* (Turko et al., 2016) were devoted to the impact of this reform on the bibliometric performance by RAS and HES, and on the collaboration among these two science bodies. The paper by Turko et al. (2016) was based on statistics from the Russian Index of Science Citation (RISC) and from Scopus. The time frame was 2010-2014.

The Russian government assigned for *Project 5-100* implementation 44 billion Rub. (around 730 million US$) for the time period 2013-2016. After two tiers of competition 14 universities were selected; in a later phase, one was added. Each year, all universities were divided for 3 groups according to their results. Each university belonging to the first group receives about 960 MLN Rub.; universities in the second group receive 450-540 MLN Rub each. Finally, each university of the third group obtains about 100 MLN Rub.

On 10 July 2015 the Government announced a new competition among universities to become more competitive among leading universities in the world included in the *Program 5-100*. On 23-24 October in the same year in Vladivostok, the International Council had the final meeting and selected six universities. These institutions were added to the list of the Top 15 universities already assigned to *Project 5-100.*

On 23.05.2016. D.Medvedev - the Prime minister of the Russian Federation signed the allocation of about 11 Billion Rub. (around 187 Million US$) to 21 universities. This input forms an additional budget beside the annual budget of each university assigned by the Ministry of Education and Science in 2017. The amount of government funding is changing yearly and depends on university performance and on the capability of elaborating and implementing competitiveness required by the Project. In 2018 according to decision of expert counsel of the Ministry of Higher Education and Science (MHES) of RF, all universities in the *Project 5-100* were divided into three groups, receiving 780 million Rub. (around 13 million US$), 480 million Rub. (around 8 million US$) and 100 million Rub. (around 17 million US$), respectively (5top100, n.d.).

On Sept.1.2017 in the framework of the *Project 5-100* a new project was announced, named Universities as a Drivers of Region's Development". 121 universities sent applications to participate in the competition as a center of innovation and social development in the region. Two independent peer-review councils were set up. 51 Universities were selected; among them were 10 universities-participants of *Project 5-100*.

## 3. Data collection

### *3.1 Scopus*

Scopus data were extracted manually from the online version (Scopus.com). Unless indicated otherwise, the data were collected in October 2017. Recently, Elsevier changed the output screens of Scopus.com, making it virtually impossible to extract frequency tables of the major database fields by means of screen-scraping or file downloading. This means that it is hardly impossible nowadays to collect useful bibliometric data tables from Scopus.com. As a consequence, in the current study the Scopus publication counts by country, year, discipline, document or source type and publication language had to be inserted manually one-by-one into a datasheet for further statistical analysis.

According to the Scopus website, in the selection of sources a Content Selection and Advisory Board (CSAB) plays an important role. The CSAB is an international group of scientists, researchers and librarians who represent the major scientific disciplines. Its board members are responsible for reviewing all titles that are suggested to Scopus (CSAB, n.d.). Despite the Board's efforts, Elsevier decided in January 2018 to discontinue 424 journals indexed in Scopus, because of "publication concerns". The effect of this decision is further analysed in Section 4.2 below.

The following document types were included in the counts: articles (abbreviation: ar), reviews (re) and conference papers (cp). In many bibliometric studies, articles and reviews are considered as the most important types of journal publications (e.g., CWTS, n.d.). A recent study has detected in Scopus duplicate records. Documents were included twice, with document type article and proceedings paper, respectively (Franceschini, Maisano & Mastrogiacomo, 2016). But the fraction of these cases was low: It is assumed that they do not substantially affect the results.

Elsevier sells biblometric information derived from Scopus in its ***SciVal*** products. In the current study ***SciVal*** was not used. A detailed analysis in the online version of Scopus by data field – especially by source (journal or conference proceedings volume) – is hampered by the fact that the online version of Scopus gives for any document set and any data field at most 160 entries. As a result, the maximum number of sources for which in a specific year publication counts can be generated in an online analysis of document set is 160.

### *3.2 WoS*

The publication counts from the Web of Science Core Collection were derived from the following databases: The Science Citation Index – Expanded (SCI-E), Social Science Citation Index (SSCI), Arts & Humanities Citation Index (AHCI), Conference Proceeding Citation Index-Science, Conference Proceeding Citation Index-Social Sciences & Humanities, Book Citation Index-Science, Book Citation Index & Social Sciences & Humanities. The total collection of these databases will be indicated as WoS throughout this paper. Sources in the WoS are selected using a combination of quantitative indicators and expert review. Clarivate Analytics sells also a bibliometric information product named ***InCites***, derived from the WoS. Most of the counts presented in the current paper are extracted with permission from ***InCites***. Unless indicated otherwise, the data were collected in October 2017.

The following document types were included in the counts: articles (abbreviation: ART), reviews (REV) and proceedings papers (PROC). For the time period 2000-2016, approximately 30,000 documents have both article and proceedings paper as document type. These multiple assignments reflect the overlap between the three WoS journal-based databases (SCI-E, SSCI, A&HCI) and the two proceedings databases. It is assumed that the conclusions from the study are not affected by this issue of double counting.

### *3.3 Other relevant datasets*

The analyses presented in this paper are based on Clarivate Analytics' Web of Science (WoS) and Elsevier's Scopus. Two other databases should be mentioned, with acronyms RISC and RSCI-C. RISC stands for the Russian Index of Science Citation, owned by a company named E.Library (RISC, n.d.). There is evidence that until the end of 2016, RISC did not have any selection process of sources, and indexed many types of Russian publications (Khantemirov, 2014). The acronym RSCI-C indicates the Russian Science Citation Index at Clarivate, a database with Russian literature created by E.Library together with Clarivate Analytics. It is a National Index, similar to Clarivate's Korean, Chinese and Latino Citation Index. This group of four indexes is included in the Web of Science Platform, but their sources are not processed for the Web of Science Core Collection or Current Contents. They are *not* analysed in the current paper.

# 4. Results

The time period taken into account in the analyses presented below is 2006-2016. Since major changes occurred during the last four years of this time period, this section also presents statistics with respect to the time period 2012-1016. Trends are characterized by compound annual growth rates (CAGR) in the annual number of published articles. In order to indicate the level of the absolute numbers, the tables also present the absolute number of publications in the final year 2016.

## *4.1 Counts per document type*

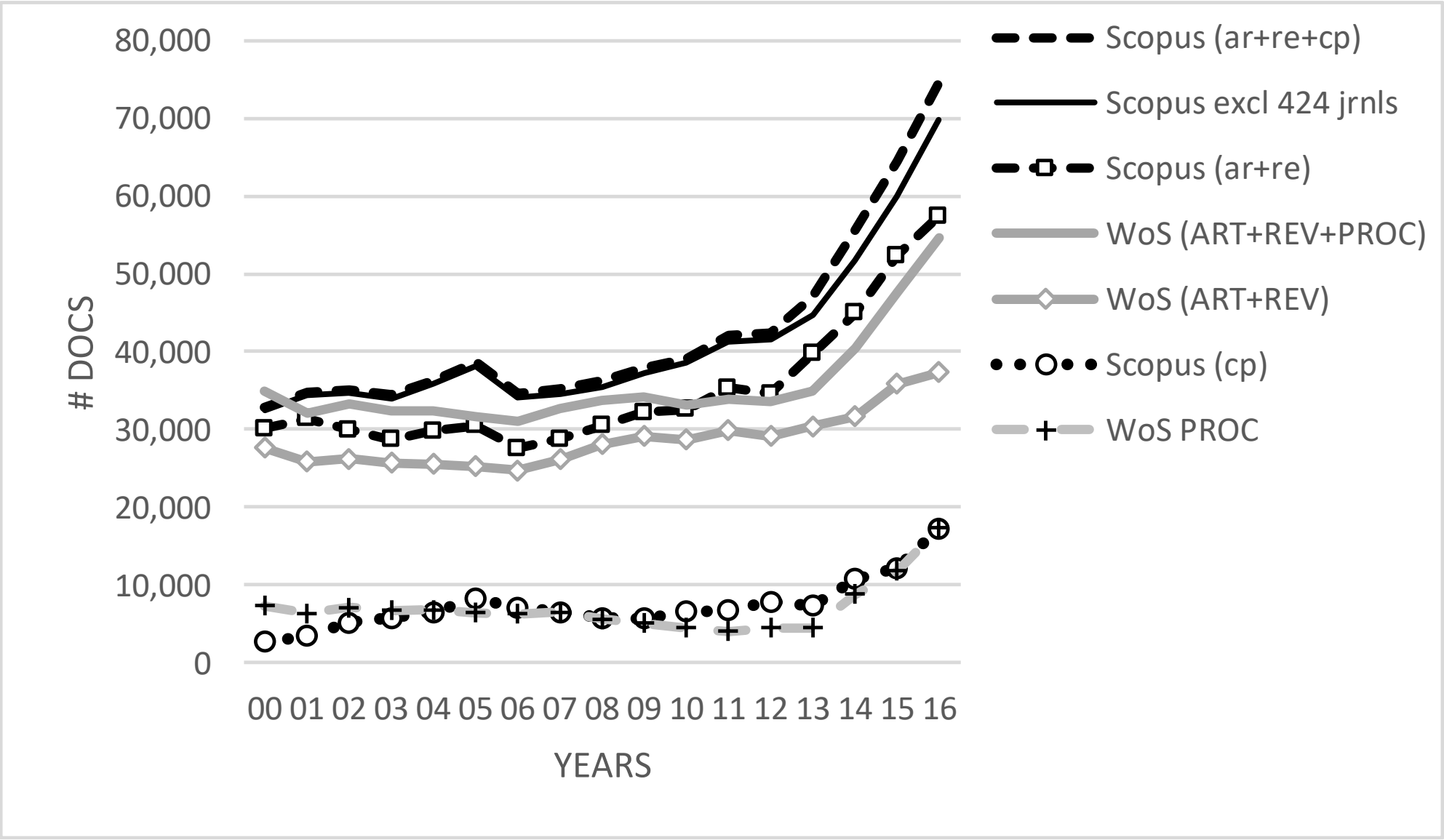

Figure 1: Number of documents from Russian institutions during 2000-2016 per year and per document type. Scopus related curves are in black, WoS-related curves in grey. The curve labelled "Scopus excl 424 jrnls" is discussed in Section 4.2 below.

Figure 1 displays the numbers of documents indexed in Scopus and WoS during 2000-2016, disaggregated by document type. It shows during the time period 2012-2016 approximately an exponential growth in the total number of documents (articles, reviews and conference papers) from Russia indexed in Scopus. In addition, it shows that during this time period also WoS revealed an exponential increase in the total number of documents from Russia, at about the same rate as that observed in Scopus.

Table 1: Compound annual growth rates (CAGR) in the number of documents from Russia by document type

| Scopus (Russian Federation) | | | | WoS (Russia) | | | |
|---|---|---|---|---|---|---|---|
| Document type | Count in 2016 | CAGR 2006-2016 | CAGR 2012-2016 | Document type | Count in 2016 | CAGR 2006-2016 | CAGR 2012-2016 |
| ar+re | 57,458 | 7.6 % | 13.5 % | ART+REV | 37,356 | 4.2 % | 6.4 % |
| cp | 17,239 | 9.3 % | 21.9 % | PROC | 17,313 | 10.7 % | 40.2 % |
| ar+re+cp | 74,697 | 8.0 % | 15.2 % | ART+REV+PROC | 54,669 | 5.9 % | 12.9 % |

Legend to Table 1: ar, re ,cp indicate articles, reviews and conference papers in Scopus. ART, REV, PROC indicate these three types in WoS.

These observations are consistent with the compound annual growth rates (CAGR) in the various types of documents from Russia indexed in Scopus and WoS, presented in Table 1. During 2012-2016, CAGR in the total number of documents from Russia is in Scopus slightly higher than that in WoS (15.2 versus 12.9). But in Scopus the number of journal articles and reviews increased much faster than that in WoS (13.5 versus 6.4), and the number of proceedings papers slower (21.9 versus 40.2). Section 4.4 provides more information on the role of proceedings papers in the two databases.

### *4.2 Recent changes in Scopus coverage*

On 28 January 2018 Elsevier published a list of 424 Scopus source journals that will be discontinued (Elsevier, n.d.). This means that the journals will not be indexed in Scopus anymore, and their backlog will be deleted from de Scopus.com database. In most cases Elsevier indicated "publication concerns", and, less often, "metrics" as the reason for discontinuation. It seems plausible to assume that these concerns are based on evidence that these journals are potentially or actually predatory (see for instance Savina & Sterligov, 2016). Predatory open-access publishing is "an exploitative open-access publishing business model that involves charging publication fees to authors without providing the editorial and publishing services associated with legitimate journals (open access or not)" ("Predatory Open Access", n.d.). Jeffrey Beall has created a list of "potential, possible, or probable predatory scholarly open-access journals" (Beall, n.d.).

Table 2: The effect of Scopus journals discontinued in 2018 upon counts of documents from Russia

| 424 discontinued journals included | | | | 424 Discontinued journals not included* | | |
|---|---|---|---|---|---|---|
| Document type | Count in 2016 | CAGR 2006-2016 | CAGR 2012-2016 | Count in 2016 | CAGR 2006-2016 | CAGR 2012-2016 |
| ar+re | 57,458 | 7.6 % | 13.5 % | 52,544 | 6.9 % | 11.7 % |
| cp | 17,239 | 9.3 % | 21.9 % | 17,239 | 9.3 % | 22.2 % |
| ar+re+cp | 74,697 | 8.0 % | 15.2 % | 69,782 | 7.4 % | 13.8 % |

*Data was collected in February 2018.

Table 2 shows that the effect of documents published in the set of the journals discontinued in January 2018 upon the counts of documents from Russian institutions is substantial, but that the overall conclusions are not affected. This is illustrated in Figure 1 in Section 4.1 that shows that even if the documents from Russia published in the 424 set are deleted, the growth of the research output from Russian institutions is still exponential.

It must be noted that of the 29,271 articles, reviews and conference papers published from Russian institutions in the set of 424 discontinued journals 99.5 percent is written in English, and about 0.4 per cent in Russian. It can be concluded that the outcomes of the analysis per publication language presented in the next section is not affected by the discontinuation of these 424 journals.

## 4.3 Counts per publication language

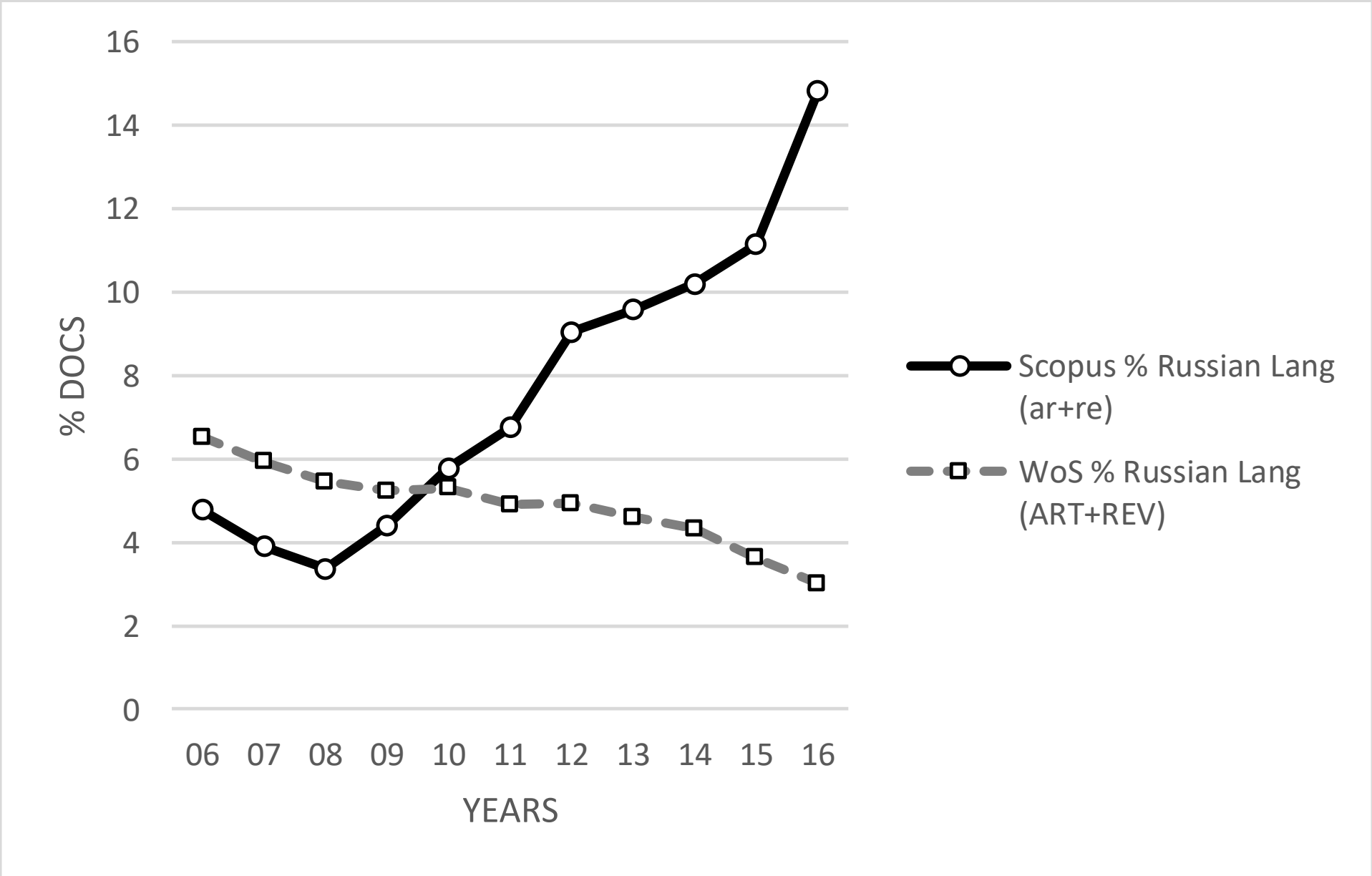

Figure 2: Percentage of documents from Russian institutions by publication language

Table 3. Number of documents (articles and reviews) from Russian institutions published in Russian language

| Database | Publication language | Count in 2016 | CAGR 2006-2016 | CAGR 2012-2016 |
|---|---|---|---|---|
| Scopus | Russian | 8,513 | 20.5 % | 28.5 % |
| | Non-Russian | 48,945 | 6.4 % | 11.7 % |
| WoS | Russian | 1,124 | -3.5 % | -6.0 % |
| | Non-Russian | 36,232 | 4.6 % | 7.0 % |

Legend to Table 3. CAGR: Compound Annual Growth Rate. Non-Russian language is in almost all cases English.

The increase in number of documents from Russian institutions indexed for Scopus is to a considerable extent due to an increase in the number of documents using Russian as *publication language*. The CAGR during 2006-2016 for documents from Russian institutions published in non-Russian languages (mainly English) is for Scopus about 40 per cent higher than it is for Wos (6.4 against 4.6), while according to Table 1, for all articles and reviews, regardless their publication language) it is as much as 81 per cent higher (7.6 versus 4.7)

*Two* analyses on *Russian language journals* were conducted. The *first* manually compared for the year 2016 the list of the 25 Russian language journals with the largest number of publications in Scopus with the same top list for WoS. They had only one journal in common, namely *Terapevticheskii Arkhiv*. This finding suggests that there are substantial differences in the coverage of Russian language journals between Scopus and WoS in 2016.

A *second* analysis focused on the Russian journals processed for *Scopus* in the year 2016 but *not* in 2012. These are the newly covered journals that are responsible for the large CAGR of Russian language journal in Scopus during 2012-2016. It was examined whether these journals are processed by VINITI for the Abstract Journal (*Referativnyi Zhurnal*). The base assumption was that the inclusion of a journal by VINITI provides an indication of its significance in its subject field. The Abstracts Journal is made up of reviews and abstracts of various published materials in a series of research areas. From a consultation of experts at VINITI it was concluded that the Abstract Journal indexes in principle all journals covered by WoS or Scopus, but that it does not cover *social sciences* and *humanities*, and that it has only partial coverage of the medical literature. In fact, the journals *not* indexed in the Abstract Journal are all in *social sciences* and *humanities*.

### *4.4 Analyses of conference proceedings documents*

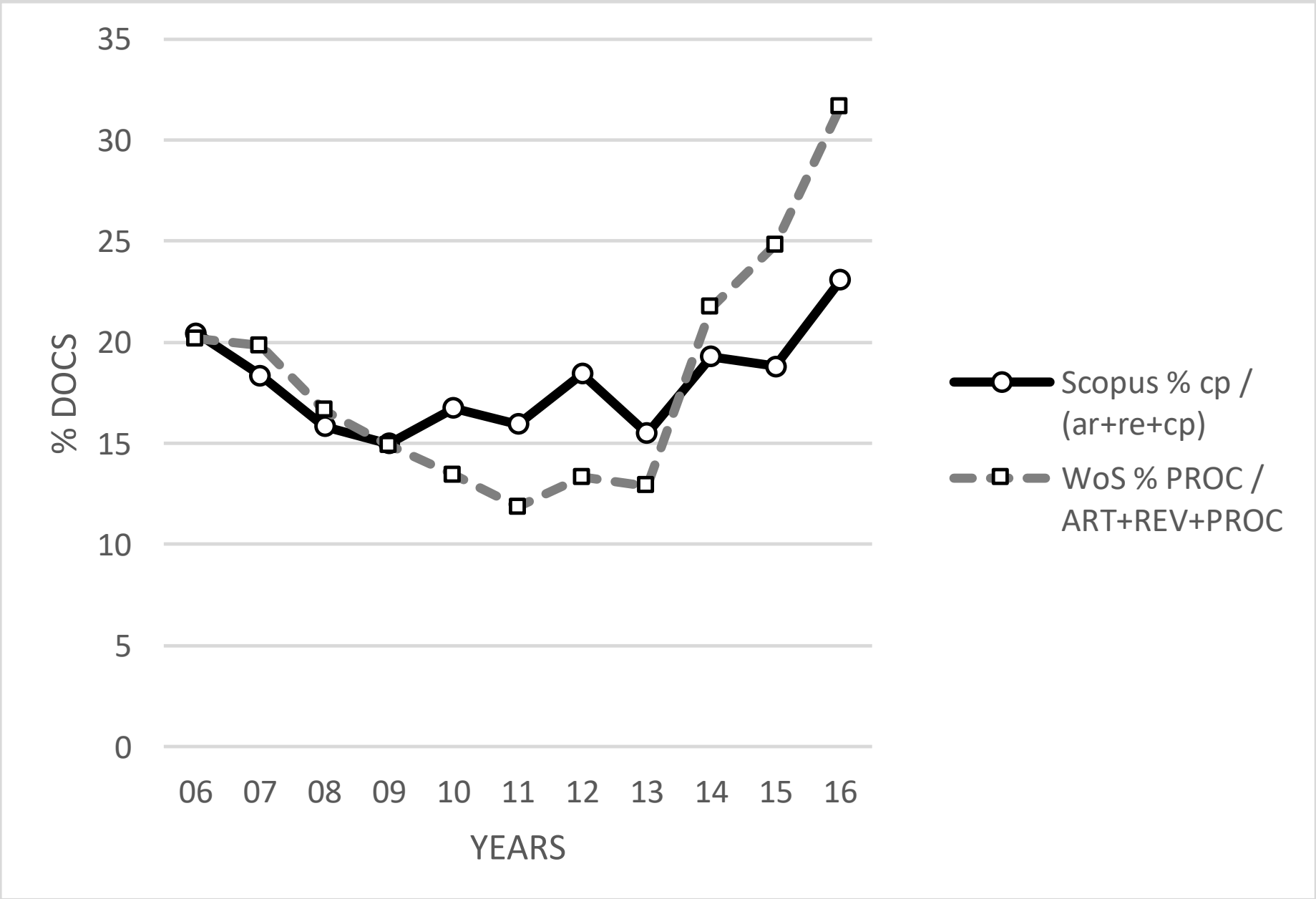

Figure 3. Percentage of conference proceedings documents (cp in Scopus, PROC in WoS) from Russian institutions. Percentages are calculated relative to the total number of indexed articles (ar, ART), reviews (re, REV) and conference proceedings papers (cp, PROC) from Russia.

Figure 3 reveals that the percentage of proceedings papers *from Russian institutions* in WoS declined during 2007-2011 from 20 to 12 per cent, followed by a sharp increase during 2014-2016. In 2016, the percentage of conference papers from Russian is 23 for Scopus against 32 for WoS.

To obtain more insight into the coverage of conference proceedings, lists of the 25 proceedings volumes with the largest number of documents from Russian institutions in 2016 were compiled for each of the two databases, and compared with one another. These lists are included in Table A1 in the Appendix. Table 4 summarizes the main outcomes. Two types of proceedings volumes are distinguished: *one-off volumes* relate to one particular conference; all its papers are presented at the same conference and published in the same year. A *conference series* publishes proceedings of a collection of conferences, for instance, volumes of subsequent annual conferences, or a set of different conferences covering a particular scientific-scholarly field.

Table 4. Publication counts for the 25 proceedings titles in Scopus and WoS with the largest number of papers from Russian institutions in 2016.

| Indicator | Scopus | WoS |
|---|---|---|
| Total conf. proceedings papers from Russian institutions in 2016 | 17,239 | 17,313 |
| Number (%) of papers in list of top 25 volumes in terms of number of papers from Russia | 11,236 (65 %) | 11,759 (68 %) |
| Number (%) papers of Top 3 volumes | 4,367 (25 %) | 3,709 (21%) |
| Overlap between Scopus and WoS top 25 lists in terms of number of papers from Russia | 15 volumes (60 %, 10 series, 5 one-off) | |

Table 4 provides evidence that in 2016 strong similarities exist between Scopus and WoS as regards the coverage of conference proceedings, not only in terms of absolute numbers of proceedings papers indexed, but also in terms of the degree of overlap of titles covered. A secondary analysis showed that the growth of conference papers from Russian institutions is unevenly distributed among institutions: four institution account for 20 per cent of the growth in Scopus-based counts.

An additional analysis relates to *citations*. For each volume in the Top 25 lists of 2016 conference proceedings in Scopus and WoS the following data was collected:

a) The total number of proceedings papers published during 2012-2016;
b) The number and percentage of papers from Russian institutions;
c) The total number of documents citing a proceedings volume and published up until January 2018.
d) The percentage of citing articles from Russia, i.e., published by Russian institutions.

Table 5. Citation counts for the 25 proceedings titles in Scopus and WoS with the largest number of papers from Russian institutions in 2016.

| Indicators | Scopus | | | | WoS | | | |
|---|---|---|---|---|---|---|---|---|
| | One-off | | Series | | One-off | | Series | |
| | Range | Median | Range | Median | Range | Median | Range | Median |
| % Russian Docs | 80-99% | 96% | 1-26% | 9% | 75-99% | 97% | 2-30% | 14% |
| % Citing Russian Docs | 82-100% | 90% | 55-98% | 79% | 75-100% | 91% | 54-100% | 81% |
| Citing non-Russian Docs /Russian Doc | 0.00-0.12 | 0.01 | 0.01-1.03 | 0.10 | 0.00-0.07 | 0.01 | 0.00-0.46 | 0.10 |

The percentage of citing documents from Russia is interpreted as a measure of the *international orientation* of a conference proceeding. The same is true for the ratio of the number of citing documents without authors from Russia and the number of papers published from Russia. The latter ratio can be conceived as a proxy of an impact factor of a proceedings title, in the calculation of which so called country self-citations are not included. These data are also presented in Table A1. Table 5 gives a statistical summary. The following conclusions could be drawn.

The outcomes for Scopus and WoS are statistically similar. As regards the *one-off* conferences, the observed large percentages of papers from Russian institutions suggest that these conferences were held in Russia and were mainly attended by Russian participants. The citation impact of these volumes on the literature published by researchers outside Russia is very low.

As regards the conference *series*, the percentage of papers from Russia tends to be much lower than those for one-off volumes. A more detailed analysis should reveal whether all covered conferences in a series showed such low percentages, or whether those related to conferences attended by Russian researchers were much higher, and similar to those of the one-off conferences on the list. In any case, the citation impact of the Russian papers outside Russia is again very low, although somewhat higher than that of the papers published in one-off volumes. This is true both for Scopus and for Wos.

## *4.5 Analysis by discipline*

Table 6. Compound annual growth rates of the number of documents from Russia (articles and reviews) per discipline

| Scopus | | | | | WoS (ESI classification) | | | | |
|---|---|---|---|---|---|---|---|---|---|
| Discipline | Nr Publ in 2016 | % Publ in 2016 | CAGR 2006-2016 | CAGR 2012-2016 | Discipline | Nr Publ in 2016 | %Publ in 2016 | CAGR 2006-2016 | CAGR 2012-2016 |
| Arts & Humanities | 2,108 | 2.2 % | 33.8 % | 38.9 % | See Social Sciences | | | | |
| Agricultural sciences | 3,771 | 3.9 % | 8.9 % | 11.5 % | Agricultural Sciences | 256 | 0.7 % | 1.5 % | 9.4 % |
| | | | | | Plant & Animal Science | 1,209 | 3.4 % | 7.3 % | 7.5 % |
| Biochemistry, Genetics & Molecular Biology | 5,895 | 6.2 % | 6.1 % | 11.0 % | Biology & Biochemistry | 1,396 | 3.9 % | 3.0 % | 7.0 % |
| | | | | | Molecular Biology & Genetics | 868 | 2.4 % | 4.7 % | 7.7 % |
| Business, Management , Accounting | 1,368 | 1.4 % | 34.1 % | 43.5 % | See Economics & Business | | | | |
| Chemistry | 8,751 | 9.1 % | 4.7 % | 9.4 % | Chemistry | 7,680 | 21.3 % | 2.9 % | 7.8 % |
| Chemical Engineering | 3,436 | 3.6 % | 9.1 % | 13.4 % | | | | | |
| Computer Science | 2,055 | 2.1 % | 13.6 % | 25.3 % | Computer Science | 558 | 1.5 % | 5.0 % | 12.8 % |
| Decision Sciences | 199 | 0.2 % | 8.6 % | 9.2 % | | | | | |
| Dentistry | 3 | 0.0 % | 4.1 % | -6.9 % | | - | | | |
| Earth & Planetary Sciences | 5,026 | 5.3 % | 5.0 % | 9.2 % | Geosciences | 2,453 | 6.8 % | 3.2 % | 4.3 % |
| | | | | | Space Science | 1,135 | 3.1 % | 3.7 % | 5.4 % |
| Economics, Econometrics & Finance | 1,595 | 1.7 % | 39.3 % | 53.3 % | Economics & Business | 125 | 0.3 % | 13.9 % | 12.9 % |
| Energy | 2,254 | 2.4 % | 6.3 % | 10.6 % | | - | | | |
| Engineering | 9,160 | 9.6 % | 12.0 % | 19.2 % | Engineering | 2,318 | 6.4 % | 6.7 % | 10.2 % |
| Environment al Science | 2,797 | 2.9 % | 14.1 % | 22.7 % | Environment/ Ecology | 678 | 1.9 % | 11.4 % | 8.7 % |
| Health Professions | 648 | 0.7 % | 29.5 % | 20.8 % | | - | | | |
| Immunology & Microbiol | 1,006 | 1.1 % | 5.4 % | 8.2 % | Immunology | 137 | 0.4 % | 6.8 % | 16.6 % |
| | | | | | Microbiology | 369 | 1.0 % | 4.3 % | 5.0 % |
| Materials Science | 8,883 | 9.3 % | 7.3 % | 9.9 % | Materials Science | 2,638 | 7.3 % | 5.8 % | 11.0 % |
| Mathematics | 5,602 | 5.9 % | 6.7 % | 10.9 % | Mathematics | 1,918 | 5.3 % | 4.6 % | 3.2 % |
| Medicine | 6,199 | 6.5 % | 15.9 % | 18.1 % | Clinical Medicine | 1,646 | 4.6 % | 5.1 % | 4.8 % |
| Multidisciplinary | 868 | 0.9 % | 30.6 % | 55.1 % | Multidisciplinary | 19 | 0.1 % | 6.6 % | 17.4 % |
| Neuroscience | 651 | 0.7 % | 7.0 % | 14.6 % | Neuroscience & Behavior | 372 | 1.0 % | -0.7 % | -8.4 % |
| Nursing | 246 | 0.3 % | 33.2 % | 14.1 % | | - | | | |
| Pharmacology, Toxicology & Pharmaceut | 1,770 | 1.8 % | 12.9 % | 32.5 % | Pharmacology & Toxicology | 396 | 1.1 % | 16.9 % | 9.4 % |
| Physics & Astronomy | 15,702 | 16.4 % | 4.6 % | 8.0 % | Physics | 9,023 | 25.0 % | 2.7 % | 4.2 % |
| Psychology | 486 | 0.5 % | 12.7 % | 32.2 % | Psychiatry/ Psychology | 272 | 0.8 % | 12.7 % | 9.8 % |
| Social Sciences | 5,183 | 5.4 % | 29.4 % | 43.9 % | Social Sciences, general (incl Arts & Humanities) | 592 | 1.6 % | 7.5 % | 7.6 % |
| Veterinary Sciences | 50 | 0.1 % | 8.1 % | 16.7 % | | - | | | |

Scopus and WoS or ***InCites*** use different subject classification systems. The first five columns in Table 6 present results for Scopus, using the Scopus subject classification into 27 disciplines. The next five columns give per Scopus discipline the results for the most similar discipline from the Essential Science Indicators (ESI) classification into 22 main research areas. It must be note that, even if a Scopus discipline and an ESI area have the same name, it does not follow that their journal sets fully overlap.

For most Scopus disciplines, the difference between the Scopus Compound Annual Growth Rate (CAGR) with the CAGR of most similar WoS area are positive, indicating that Scopus numbers increase faster than WoS counts do in most disciplines. Scopus disciplines showing the largest differences in CAGR 2012-2016 are: *Economics*, *Social Science*, *Arts & Humanities*, *Pharmacology & Toxicology*, *Neuroscience*, *Environmental Science*, and *Computer Science*.

The annual growth rates of a discipline may also be affected by changes in the classification system, for instance, when journals are moved from one discipline to the other. This might for instance be the case for the category *Neuroscience & Behaviour* in WoS, that shows *negative* growth rates.

A comparison between Scopus and WoS with respect to percentage of publications from Russia assigned to a discipline, relative to the total number of assignments across all disciplines (including double counts due to multiple assignments), reveals large differences in the distribution of publications across disciplines between the two databases. Although these differences are partly due to differences in the definition of the various disciplines, they clearly show for the Russian output a relatively strong representation of Social Science & Humanities in Scopus, and of Physics and Chemistry in WoS.

### *4.6 Analysis of BRIC countries and most productive countries*

To put the results for Russia in perspective, publication counts and growth rates for this country were compared with those from other BRIC countries, and with the ten countries with the largest publication output in 2016 (excluding the two BRIC countries China and India). What the BRIC countries have in common is that they are rapidly further developing and internationalizing their research infrastructures. These countries have been the object of research in several bibliometric studies (e.g., Bornmann, Wagner, & Leydesdorff, 2015; Finardi & Buratti, 2016). Table 7 shows CAGR during 2006-2016 and 2012-2016 in the number of documents (articles and reviews) published from the various countries and indexed in Scopus and WoS. During the entire period 2006-2016, CAGR of papers from Russia is in Scopus almost twice this rate in WoS (7.6 versus 4.2 per cent).

Table 7. Annual growth rates in Scopus and WoS of the number of articles and reviews from BRIC and most productive countries

| | SCOPUS | | | | INCITES | | | |
|---|---|---|---|---|---|---|---|---|
| Country | Nr Publ 2016 | % Publ 2016 | CAGR 2006-2016 | CAGR 2012-2016 | Nr Publ 2016 | % Publ 2016 | CAGR 2006-2016 | CAGR 2012-2016 |
| *BRIC countries* | | | | | | | | |
| Brazil | 59,088 | 2.8 % | 8.2 % | 5.2 % | 47,269 | 2.9 % | 9.2 % | 4.9 % |
| China | 400,741 | 19.3 % | 9.8 % | 9.6 % | 309,441 | 18.7 % | 14.4 % | 14 % |
| India | 105,668 | 5.1 % | 10.8 % | 7.2 % | 66,738 | 4.0 % | 8.9 % | 7 % |
| Russia | 57,458 | 2.8 % | 7.6 % | 13.5 % | 37,356 | 2.3 % | 4.2 % | 6.4 % |
| *Most productive countries** | | | | | | | | |
| United States | 463,882 | 22.0 % | 2.7 % | 1.5 % | 443,325 | 26.5 % | 3.2 % | 2.0 % |
| United Kingdom | 141,709 | 6.7 % | 3.5 % | 2.9 % | 136,118 | 8.1 % | 4.3 % | 3.6 % |
| Germany | 126,803 | 6.0 % | 3.2 % | 2.3 % | 117,493 | 7.0 % | 3.8 % | 2.9 % |
| Japan | 93,327 | 4.4 % | 0.6 % | 0.2 % | 82,329 | 4.9 % | 0.5 % | 0.7 % |
| France | 89,004 | 4.2 % | 3.5 % | 2.2 % | 80,532 | 4.8 % | 3.4 % | 2.5 % |
| Italy | 81,761 | 3.9 % | 5.4 % | 4.7 % | 73,748 | 4.4 % | 5.2 % | 4.5 % |
| Canada | 77,238 | 3.7 % | 3.9 % | 2.4 % | 74,222 | 4.4 % | 4.3 % | 3.0 % |
| Australia | 73,200 | 3.5 % | 7.4 % | 6.2 % | 70,270 | 4.2 % | 8.3 % | 7.1 % |
| Spain | 72,621 | 3.4 % | 6.0 % | 3.5 % | 62,500 | 3.7 % | 6.1 % | 2.6 % |
| South Korea | 68,043 | 3.2 % | 9.5 % | 5.9 % | 60,130 | 3.6 % | 7.6 % | 4.5 % |

*Data for these 10 countries were collected in April 2018. Due to the time delay in processing, the number of documents with publication year 2016 indexed up until April 2018 is about 1.5 per cent higher than the number of 2016-documents indexed up until October 2017.

During 2012-2016 the Scopus annual growth rate is for Russia disproportionally large not only compared to that for WoS, but also compared to that of the other three BRIC countries in Scopus, and to that of the ten most productive countries. Apparently, Scopus has given a high priority to indexing Russian publication output, and WoS to publications from China. In the interpretation of these results one should bear in mind that during this time period Clarivate (formerly Thomson Reuters) created a Russian Citation Index in the Web of Science Platform. Counts derived from this database are *not* included in the results presented for WoS.

### *4.7 Analysis by journal in Scopus*

To which extent is the increase of documents from Russian institutions in Scopus journals due to an *expansion* of the journal coverage, i.e., by adding new journals to the database? Due to the limited capabilities of the online version of Scopus to collect bibliometric data, mentioned in Section 3.1, a *first* analysis by source presented in this sub-section is limited to the *50 English language* journals *with the largest number of documents from Russian institutions in 2016*. Table A2 in the Appendix presents a list of these journals. The number of 2016 documents from Russia in these journals is 13,236, which amounts to 23 per cent of the total number of the 57,458 documents from Russian institutions published in 2016. Of these 13,236 documents, 73 per cent is published in journals in which Russian researchers had published at least one paper also in 2012.

Although it must be noted that the sample analysed (the 50 journals with the largest number of documents in 2016) is biased towards voluminous journals, so that the outcome cannot be generalized, this analysis suggests that a substantial part of the increase in English language journal output from Russia is due to the fact that Russian researchers started publishing more in journals they were already familiar with in the past. But on the basis of the available data it is impossible to give a precise estimate. Table A2 shows that three journals in which Russia had zero papers in 2012, were discontinued in 2018. They published in 2016 773 Russian publications accounting for less than 6 % of the total Russian output in that year. This means that the discontinued journals had only a small affect upon the increase of the Russian output written in English.

Two journals (N5 "Physical Review B" and N 12 "Physical Review D") are published by American Physical Society. These journals have been sources of publications by the best Soviet and Russian physicists during last forty years, especially since the "Perestrojka", up until today. Table A2 shows that 35 journals (70 per cent) are published by a publisher located in Russia. These belong to the core group Soviet and Russian journals, some of which were established more than 60 years ago. Many of these journals were translated into English by a Russian company "MAIK-Nauka (Interperiodika)" or by Springer.

A second analysis relates to *Russian language* journals. In this set, the number of periodicals with publications from *Russian institutions* amounts to 58 in 2012, and about 160 in 2016. For these journals a more detailed analysis was performed. It was found that in the set of 160 Russian language journals used by Russian institutions in 2016, 117 journals (73 per cent) did not contain any documents from Russian institutions in 2012. These 117 journals accounted for about 60 per cent of the total number of documents from Russian institutions and published in Russian language journals in 2016. Although it cannot be decided whether these journals were founded after 2012 or whether they were already active in 2012 but indexed in Scopus as from 2016, from the point of view of Scopus coverage one can conclude that they are "new" journals in 2016 compared to the situation in 2012.

### *4.8 The 2.44 per cent norm for the share of Russian publication output*

Table 8 shows the trend during 2012-2016 in the percentage of documents from Russian institutions by database and type of document. It shows that it depends upon the database analysed, and the type of document included. Counting articles, reviews and proceedings papers, the percentage of Russian output exceeds in both databases the 2.44 per cent norm released by President Putin in May 2012. Leaving out proceedings papers, the Russian share in 2016 does exceed the norm in Scopus, but *not* in WoS. Deleting the 424 discontinued journals in Scopus, this share further declines but remains still above the norm, but if papers in Russian language are discounted as well, it declines to 2.18 per cent, substantially below the 2.44 per cent norm.

Table 8. Trends 2012-2016 in percentage of documents from Russia by database and document type

| | Scopus ar+re+cp | Scopus ar+re | Scopus excl. 424 jrnls ar+re | Scopus excl 424 jrnls excl Russian Lang ar+re | WoS ART+REV+PROC | WoS ART+REV |
|---|---|---|---|---|---|---|
| 2012 | 1.79 % | 1.87 % | 1.92 % | 1.75 % | 1.87 % | 1.95 % |
| 2013 | 1.91 % | 2.01 % | 2.03 % | 1.82 % | 1.86 % | 1.94 % |
| 2014 | 2.19 % | 2.18 % | 2.16 % | 1.93 % | 2.08 % | 1.97 % |
| 2015 | 2.60 % | 2.52 % | 2.37 % | 2.08 % | 2.39 % | 2.18 % |
| 2016 | 2.99 % | 2.77 % | 2.60 % | 2.18 % | 2.72 % | 2.25 % |

### *4.9 Trends in domestic and international collaboration*

Figure 4 shows trends were analyzed in domestic collaboration (DC) and international collaboration (IC) based on Russian publications covered by Scopus and WoS. It reveals a significant growth in the share of DC in both databases. A similar trend by observed in the analysis of WoS data by Mindeli et. all (2016).

There is a notable difference in the pattern of IC among the two databases. Scopus reveals a decline in IC share during the studied period. This declining trend can be partly attributed to the growth of the Scopus coverage of Russian journals. Probably, these journals do not belong to the group of more prestigeous journals in which foreign coauthors prefer to publish their research results.

In WoS the IC share declined during 2006 till 2011, increased during 2011-2013, but then declined again. The latter declining trend could be related to the government reform of the main research body in Russia – the Russian Academy of Sciences.

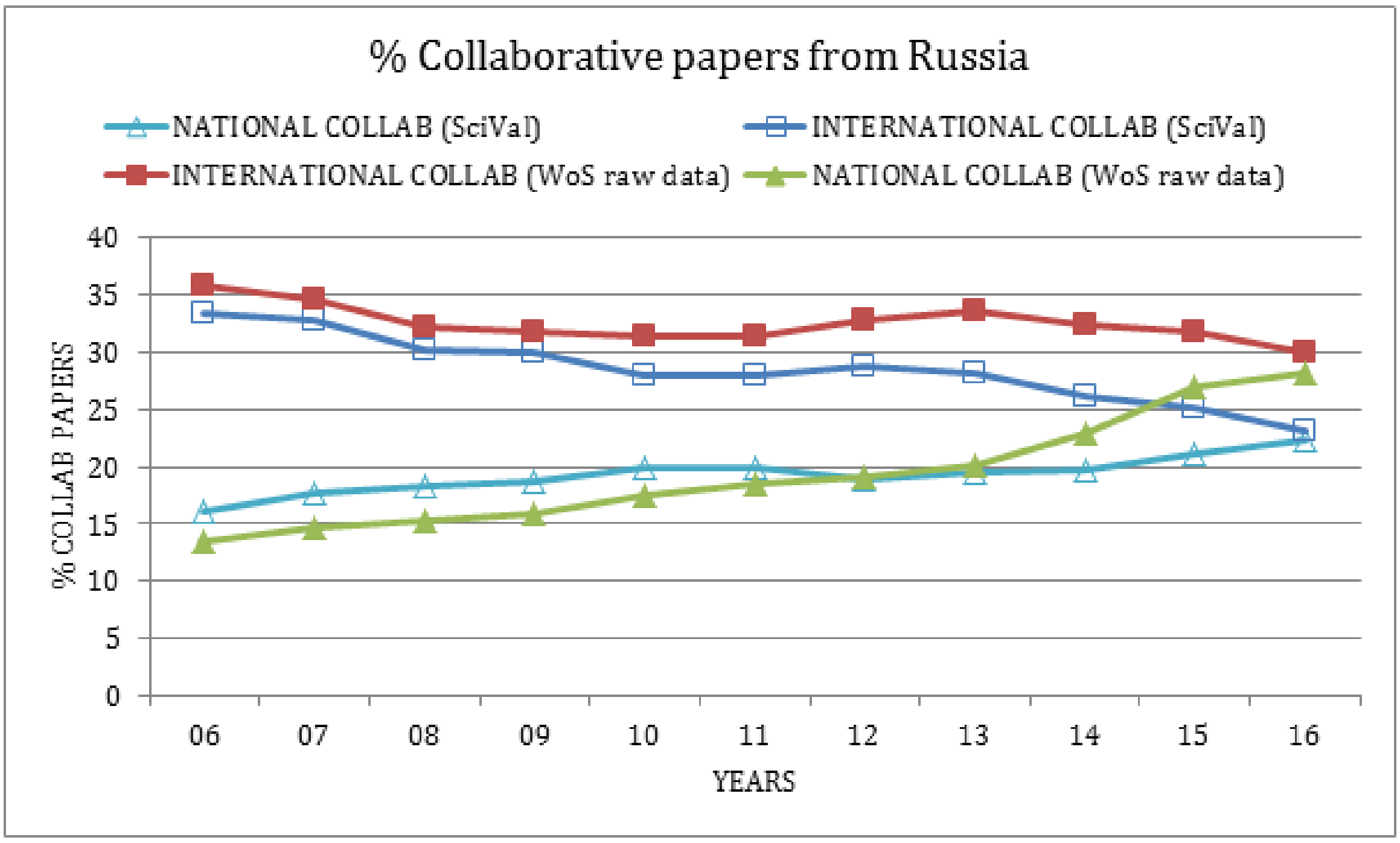

Figure 4. Trends in Russian domestic and international collaboration. Data were exported from analytical tool SciVal (Scopus) on 16 March 2018. The WoS raw records were downloaded via Web of Science- Web Services Premium Clarivate Analytics, the query CU=Russia (databases: SCI, SSCI, AHCI, ISTP, ISSHP, BSCI, BHCI). The set of records (art+rev+proc) during 2006 -2014 was downloaded on 5.09.2016. of September 2016, and the set of records from 2015 to 2016 was download on 5.04.2018.

## 5. Discussion and conclusions

### *5.1 Conclusions*

The analyses presented in Section 4 allow for the following conclusions. A *first* conclusion is that the calculation of numbers and growth rates of documents published from Russian institutions very much depends upon *the database* that is used. This finding points towards a serious problem from a user perspective: as both WoS and Scopus strive to be a standard in bibliometric research assessment – the former through its product ***InCites*** and the latter via ***SciVal*** –, a user may be confronted with substantial differences in outcomes between the two. Without additional information about the coverage of these databases he is not able to assess which outcome is the most valid. The current study on Russian publication output provides a clear illustration of this problem. The share of papers from Russia indexed in Scopus is higher in Scopus than it is in WoS. A *second* conclusion of holds that, if one uses one single database, outcomes may be affected by *changes* in database coverage. This is also clearly illustrated in the current study for Russian output. The increase in share of Russian papers during 2012-2016 depends in both databases upon the expansion of their proceedings coverage, and, in the case of Scopus, also upon the inclusion of more Russian language journals.

The initial finding presented at the NEICON conference of an exponential increase during 2012-2016 in the number of documents from Russia indexed in *Scopus* is confirmed. But the current analysis reveals a similar exponential pattern also in *WoS*. The compound annual growth rates (CAGR) are statistically similar in both databases (15.2 % for Scopus against 12.9 % for WoS during 2012-2016). A key difference is that in Scopus the number of journal articles and reviews increased faster than that in WoS, and the number of *proceedings papers* slower. The discussion below focuses first on journal publications, and deals next with conference proceedings papers.

As regards *journal publications* (articles and reviews), during 2006-2016 the Compound Annual Growth Rate (CAGR) of the number of articles and reviews from Russia indexed for Scopus is about *80 per cent higher* than this rate found in WoS. Evidence was found that during 2012-2016 the CAGR in Scopus of published journal papers is for Russia disproportionally large not only compared to that of the other three BRIC countries in Scopus, but also compared to this rate in WoS. The primacy of Scopus over WoS in terms of CAGR of number of papers is observed in most disciplines, especially in *Economics*, *Social Science*, *Arts & Humanities*, and also in *Pharmacology & Toxicology*, *Neuroscience*, *Environmental Science*, *and Computer Science.*

The large increase in the number of documents from Russian institutions indexed for Scopus is to a considerable extent due to the fact that Scopus increased the indexing of papers using Russian as *publication language.* While in WoS the percentage of papers in Russian language journals declined during 2006-2016 from 6.5 to 3.0 per cent, in Scopus it increased from 4.8 to 14.8 per cent. Hardly any overlap was found between Scopus and WoS in their sets of most important Russian journals in terms of number of papers published.

The current study found that Scopus has substantially *expanded the coverage* of Russian language journals. About three quarters of the Russian language journals covered in Scopus in 2016 were not indexed in 2012. This expansion explains a substantial part of the exponential increase in the annual number of documents published from Russian institutions during 2012-2016. On the other hand, it was found that another part of the increase in journal output from Russia during 2012-2016 is not due to an expansion of the journal coverage, but rather to the fact that Russian researchers started publishing more in journals they were already familiar with and that they used in 2012.

This latter finding provides evidence that Russian institutions increased their publication output in internationally oriented journals. But the extent to which these 'new' publications displaced articles that Russian scientists in earlier years tended to publish in national journals not covered by Scopus or WoS, cannot be assessed with the data presented in Section 4. An analysis of the Russian output in the Russian Index of Science Citation (RISC) or Russian Science Citation Index (RSCI-C) could reveal relevant insights into displacement but falls beyond the scope of the current paper.

As regards the *proceedings papers*, CAGR during 2006-2016 of the number of *conference proceedings articles* in Scopus is statistically similar to that in WoS; both are around 10 per cent. In WoS the percentage of proceedings articles declined during the first half of this time period, and strongly increased during 2012-2016, catching up with Scopus. In 2016, the percentage of conference papers is 23 for Scopus against 32 for WoS. There is a substantial overlap between Scopus and WoS in proceedings titles covered, especially *conference series.* The 'big' series tend to be covered in both.

The study of the 25 proceedings titles with the largest number of papers from Russia in 2016 provided evidence that the one-off proceedings volumes published almost exclusively papers from Russia and relate to conferences organized in Russia itself. The citation impact generated up to date by these volumes on the literature published by researchers outside Russia tends to be almost zero. Although the percentage of Russian papers in conference series is much lower than it is in one-off proceedings, the citation impact is again rather low. These conclusions are valid both for Scopus and for WoS.

From a bibliometric point of view, it is questionable whether a positive trend in the number of documents from Russian institutions published in Russian language journals indexed in Scopus reflects a genuine internationalization of Russian research. The same question should be raised as regards the strong increase in the number of conference papers by Russian researchers indexed in the two databases.

What the precise effects of the observed differences between database and of changes in database coverage will be upon an assessment of the performance of the Russian research system strongly depends upon the assessment methodology that is applied, and, if bibliometric indicators are to play a role, which indicators are used. If database features influence the value of multiple indicators, the effects may be in different directions.

For instance, an increase of the number of indexed publications in Russian language journals has probably a *positive* effect upon a size-independent publication output indicator. But there is strong evidence that domestic, non-English publications have a *negative* influence upon relative citation rates, comparing the citation-per-publication ratio of an institution with the world average citation rate in the subfields in which it is active (Van Leeuwen et al., 2001).

What the effect of an increase in Russian language publications and papers in proceedings of nationally oriented conferences will be on the position of Russian institutions in World University Rankings is difficult to predict without a detailed analysis. The THES and QS rankings in 2016 and 2017 use Scopus as a bibliometric data source, while the ARWU ranking is mostly based upon WoS. The effects also depend upon the way in which the Ranking producers measure publication output and citation impact, and which weights they give to these indicators.

The elementary citation analysis conducted in the current study raises another important issue. Although Table A1 in the Appendix shows that the absolute number of citations to the total collection of Russian papers in a particular proceedings title may be in the order of magnitude of several hundreds, a large fraction of these are country self-citations (i.e. , citations given in articles published by Russian authors). In fact, the values of the impact factor proxies calculated in this study are extremely low.

High percentages of author- or country self-citations, combined with low absolute citation levels may easily lead to statistical outliers in the calculation of subfield normalized, relative citation rates of the type calculated in ***SciVal*** and ***InCites***, and, hence, possibly affect university ranking systems that use these relative rates. A preliminary analysis of the citation impact of the proceedings papers of 21 Russian institutions in ***InCites*** showed that per publication year there were 1-2 institutions with a relative citation rate above 3.0, which is very high indeed, and even three cases with a rate above 5. Further research into this issue is recommended.

All in all, the results obtained in the current study provide evidence that one should be cautious when using WoS, and especially Scopus, as a measuring device of changes in research performance from an international perspective, and, hence, as a valid tool in the assessment of the key objectives of the Project 5-100.

### *5.2 Concluding remarks*

The current authors are aware that Russian research has a high international level and a great potential. The empirical findings presented in this paper does not change this view. The authors aim to contribute to developing proper tools to further demonstrate it. Being bibliometric/informetric researchers, this gives them the responsibility to critically examine the value of methods proposed by others. The current paper aims to do so. A general conclusion is that a framework for conducting critical, independent assessments of bibliographical databases and their application in research assessment is urgently needed, involving both organizational, technical and theoretical aspects (Moed, 2017). Such a framework is as of yet missing, as current assessment studies – including the one presented in the current paper – tend to be made on an ad-hoc basis, and the underlying, large scale datasets that are needed to analyse and systematically compare databases, are mostly unavailable. Adequate concordances between classification systems used in different databases – e.g., subject and document classification systems – are needed as well. In addition, dedicated data handling and analysis software is needed, and sufficient background knowledge of the ins and outs of the databases at stake, as well as the pros and cons of the use of bibliometric or informetric indicators in research assessment.

**Acknowledgement**

The authors wish to thank Dr Alexander Libkind from the All Russian Institute for Scientific &Technical Information of the Russian Academy of Sciences (VINITI) for his comments on an earlier version of this paper. They are also grateful to two anonymous referees for their valuable comments on an earlier version of our manuscript.

# Appendix

Table A1: The 25 proceedings titles in Scopus and WoS with the largest number of papers from Russian institutions in 2016.

| Title | Type* | Over-lap** | Total Publ 2012-2016 | % Russian Docs | Citing Docs (2012-Jan 2018) | % Citing Russian Docs | Non-Russian Citing Docs per Russian Doc |
|---|---|---|---|---|---|---|---|
| ***SCOPUS*** | | | | | | | |
| Lecture Notes In Computer Science Including Subseries Lecture Notes In Artificial Intelligence And Lecture Notes In Bioinformatics | 1 | Y | 98,303 | 1.5 % | 3,461 | 55.4 % | 1.03 |
| Proceedings Of SPIE The International Society For Optical Engineering | 1 | Y | 65,730 | 4.2 % | 1,485 | 81.8 % | 0.10 |
| Aip Conference Proceedings | 1 | Y | 31,026 | 8.7 % | 1019 | 87.9 % | 0.05 |
| Journal Of Physics Conference Series | 1 | Y | 26,344 | 15.9 % | 1,125 | 79.6 % | 0.05 |
| Procedia Engineering | 1 | Y | 17,897 | 8.5 % | 1,742 | 75.9 % | 0.28 |
| Ceur Workshop Proceedings | 1 | | 15,195 | 6.3 % | 506 | 79.8 % | 0.11 |
| Progress In Biomedical Optics And Imaging Proceedings Of SPIE | 1 | | 10,646 | 5.0 % | 439 | 78.4 % | 0.18 |
| Procedia Computer Science | 1 | | 8,988 | 4.5 % | 666 | 64.0 % | 0.59 |
| Iop Conference Series Materials Science And Engineering | 1 | Y | 7,792 | 25.2 % | 1,449 | 87.2 % | 0.09 |
| EPJ Web Of Conferences | 1 | Y | 7,544 | 13.2 % | 936 | 68.6 % | 0.29 |
| Matec Web Of Conferences | 1 | Y | 5,888 | 19.0 % | 669 | 97.8 % | 0.01 |
| International Multidisciplinary Scientific Geoconference Surveying Geology And Mining Ecology Management Sgem | 1 | Y | 4,654 | 12.1 % | 300 | 92.3 % | 0.04 |
| Physics Procedia | 1 | | 3,943 | 16.4 % | 941 | 58.3 % | 0.60 |
| Iop Conference Series Earth And Environmental Science | 1 | Y | 2,711 | 15.8 % | 288 | 89.2 % | 0.07 |
| International Conference Of Young Specialists On Micro Nanotechnologies And Electron Devices Edm | 2 | | 495 | 95.2 % | 322 | 82.3 % | 0.12 |
| Proceedings 2016 International Conference Laser Optics Lo 2016 | 2 | Y | 412 | 83.5 % | 19 | 84.2 % | 0.01 |
| 2016 2nd International Conference On Industrial Engineering Applications And Manufacturing Icieam 2016 Proceedings | 2 | Y | 409 | 98.5 % | 46 | 91.3 % | 0.01 |
| 2016 International Siberian Conference On Control And Communications Sibcon 2016 Proceedings | 2 | Y | 219 | 96.8 % | 149 | 88.6 % | 0.08 |
| Proceedings Of The 2016 IEEE North West Russia Section Young Researchers In Electrical And | 2 | Y | 206 | 95.6 % | 172 | 89.0 % | 0.10 |
| Electronic Engineering Conference Eiconrusnw 2016 | | | | | | | |
| 7th Eage Saint Petersburg International Conference And Exhibition Understanding The Harmony Of The Earth S Resources Through Integration Of Geosciences | 2 | | 197 | 80.7 % | 20 | 90.0 % | 0.01 |
| Proceedings Of The 19th International Conference On Soft Computing And Measurements Scm 2016 | 2 | | 172 | 98.3 % | 109 | 91.7 % | 0.05 |
| Proceedings Of 2015 International Conference On Mechanical Engineering Automation And Control Systems Meacs 2015 | 2 | | 130 | 96.2 % | 85 | 89.4 % | 0.07 |
| Geomodel 2016 18th Science And Applied Research Conference On Oil And Gas Geological Exploration And Development | 2 | | 128 | 99.2 % | 4 | 100 % | 0.00 |
| 2016 13th International Scientific Technical Conference On Actual Problems Of Electronic Instrument Engineering Apeie 2016 Proceedings | 2 | Y | 126 | 98.4 % | 16 | 100 % | 0.00 |
| 23rd Saint Petersburg International Conference On Integrated Navigation Systems Icins 2016 Proceedings | 2 | | 113 | 80.5 % | 33 | 97.0 % | 0.01 |
| ***WoS*** | | | | | | | |
| PROCEEDINGS OF SPIE (Intern.Soc. for Optic and Photonic) | 1 | Y | 78,265 | 4.1 % | 1,618 | 74.8 % | 0.13 |
| LECTURE NOTES IN COMPUTER SCIENCE | 1 | Y | 49,614 | 1.7 % | 827 | 54.1 % | 0.46 |
| AIP CONFERENCE PROCEEDINGS | 1 | Y | 35,258 | 7.9 % | 1,561 | 76.6 % | 0.13 |
| JOURNAL OF PHYSICS CONFERENCE SERIES | 1 | Y | 23,716 | 15.4 % | 3052 | 67.7 % | 0.27 |
| PROCEDIA ENGINEERING | 1 | Y | 17,414 | 8.4 % | 942 | 74.1 % | 0.17 |
| EPJ WEB OF CONFERENCES | 1 | Y | 7,281 | 13.1 % | 622 | 65.1 % | 0.23 |
| INTERNATIONAL MULTIDISCIPLINARY SCIENTIFIC GEOCONFERENCE SGEM | 1 | Y | 6,999 | 12.9 % | 126 | 88.9 % | 0.02 |
| IOP CONFERENCE SERIES MATERIALS SCIENCE AND ENGINEERING | 1 | Y | 6,902 | 22.1 % | 755 | 87.9 % | 0.06 |
| MATEC WEB OF CONFERENCES | 1 | Y | 5,783 | 18.3 % | 257 | 97.3 % | 0.01 |
| INTERNATIONAL MULTIDISCIPLINARY SCIENTIFIC CONFERENCES ON SOCIAL SCIENCES AND ARTS BULGARIA | 1 | | 3,576 | 30.3 % | 102 | 97.1 % | 0.00 |
| IOP CONFERENCE SERIES EARTH AND ENVIRONMENTAL SCIENCE | 1 | Y | 2,471 | 14.9 % | 130 | 84.6 % | 0.05 |
| SHS WEB OF CONFERENCES | 1 | | 1,649 | 18.1 % | 46 | 100 % | 0.00 |
| 2016 INTERNATIONAL CONFERENCE LASER OPTICS LO | 2 | Y | 413 | 83.1 % | 0 | . | 0.00 |
| 2016 2ND INTERNATIONAL CONFERENCE ON INDUSTRIAL ENGINEERING APPLICATIONS AND MANUFACTURING ICIEAM | 2 | | 409 | 98.3 % | 1 | 100 % | 0.00 |
| 2ND INTERNATIONAL CONFERENCE ON INDUSTRIAL ENGINEERING ICIE 2016 | 2 | Y | 377 | 97.6 % | 95 | 74.7 % | 0.07 |
| INTERNATIONAL CONFERENCE ON ACTUAL PROBLEMS OF ELECTRONIC INSTRUMENT ENGINEERING | 2 | Y | 315 | 97.8 % | 12 | 83.3 % | 0.01 |
| 22ND INTERNATIONAL SYMPOSIUM ON ATMOSPHERIC AND OCEAN OPTICS ATMOSPHERIC PHYSICS | 2 | | 266 | 98.9 % | 6 | 83.3 % | 0.00 |
| 15TH INTERNATIONAL SCIENTIFIC CONFERENCE UNDERGROUND URBANISATION AS A PREREQUISITE FOR SUSTAINABLE DEVELOPMENT | 2 | | 244 | 75.4 % | 93 | 87.1 % | 0.07 |
| ADVANCED MATERIALS WITH HIERARCHICAL STRUCTURE FOR NEW TECHNOLOGIES AND RELIABLE STRUCTURES 2016 | 2 | | 239 | 97.9 % | 50 | 94.0 % | 0.01 |
| 2016 11TH INTERNATIONAL FORUM ON STRATEGIC TECHNOLOGY IFOST PTS 1 AND 2 | 2 | | 224 | 86.2 % | 0 | . | 0.00 |
| IEEE INTERNATIONAL SIBERIAN CONFERENCE ON CONTROL AND COMMUNICATIONS | 2 | | 219 | 97.3 % | 2 | 100 % | 0.00 |
| 2016 INTERNATIONAL SIBERIAN CONFERENCE ON CONTROL AND COMMUNICATIONS SIBCON | 2 | Y | 219 | 97.3 % | 2 | 100 % | 0.00 |
| 3RD INTERNATIONAL SCHOOL AND CONFERENCE ON OPTOELECTRONICS PHOTONICS ENGINEERING AND NANOSTRUCTURES SAINT PETERSBURG OPEN 2016 | 2 | | 209 | 93.8 % | 53 | 81.1 % | 0.05 |
| PROCEEDINGS OF THE 2016 IEEE NORTH WEST RUSSIA SECTION YOUNG RESEARCHERS IN ELECTRICAL AND ELECTRONIC ENGINEERING CONFERENCE ELCONRUSNW | 2 | Y | 206 | 95.6 % | 90 | 91.1 % | 0.04 |
| XXXI INTERNATIONAL CONFERENCE ON EQUATIONS OF STATE FOR MATTER ELBRUS 2016 | 2 | | 205 | 96.6 % | 33 | 75.8 % | 0.04 |

Legend to Table A1. *: Type: 1: Conference Series. 2: One-off volume. ** Overlap: Y: Title is included in the list of the second database.

Table A2. Top 50 English language journals with the largest number of papers from Russia in 2016

| Rank | Source title | N Docs 2016 | N Docs 2012 | CAGR 2012-2016 | Publisher's Country | Comments |
|---|---|---|---|---|---|---|
| 1 | International Journal Of Environmental And Science Education | 653 | 0 | . | Netherlands | |
| 2 | Research Journal Of Pharmaceutical Biological And Chemical Sciences | 527 | 0 | . | India | |
| 3 | Key Engineering Materials | 465 | 1 | 364.4 % | Switzerland | |
| 4 | Russian Chemical Bulletin | 405 | 275 | 10.2 % | Russia | |
| 5 | Physical Review B | 388 | 355 | 2.2 % | USA | |
| 6 | Indian Journal Of Science And Technology | 377 | 0 | . | India | |
| 7 | Physics Of The Solid State | 372 | 350 | 1.5 % | Russia | |
| 8 | Bulletin Of Experimental Biology And Medicine | 337 | 418 | -5.2 % | Russia | |
| 9 | Bulletin Of The Russian Academy Of Sciences Physics | 331 | 99 | 35.2 % | Russia | |
| 10 | Materials Science Forum | 330 | 0 | . | Switzerland | |
| 11 | International Journal Of Pharmacy And Technology | 328 | 0 | . | India | Discon-tinued |
| 12 | Physical Review D | 322 | 314 | 0.6 % | USA | |
| 13 | Journal Of Mathematical Sciences United States | 319 | 147 | 21.4 % | Russia | |
| 14 | Russian Journal Of General Chemistry | 301 | 251 | 4.6 % | Russia | |
| 15 | Technical Physics Letters | 297 | 269 | 2.5 % | Russia | |
| 16 | Doklady Earth Sciences | 295 | 219 | 7.7 % | Russia | |
| 17 | Technical Physics | 292 | 258 | 3.1 % | Russia | |
| 18 | Russian Journal Of Organic Chemistry | 281 | 215 | 6.9 % | Russia | |
| 19 | International Journal Of Applied Engineering Research | 280 | 0 | . | India | |
| 20 | JETP Letters | 280 | 247 | 3.2 % | Russia | |
| 21 | Russian Journal Of Physical Chemistry A | 279 | 246 | 3.2 % | Russia | |
| 22 | International Review Of Management And Marketing | 267 | . | . | Turkey | Discon-tinued |
| 23 | Semiconductors | 263 | 209 | 5.9 % | Russia | |
| 24 | Scientific Reports | 260 | 14 | 107.6 % | Great Britain | |
| 25 | Russian Engineering Research | 236 | 173 | 8.1 % | Russia | |
| 26 | Russian Journal Of Inorganic Chemistry | 233 | 224 | 1.0 % | Russia | |
| 27 | Optics And Spectroscopy English Translation of Optika I Spektroskopiya | 226 | 142 | 12.3 % | Russia | |
| 28 | Mathematics Education | 219 | 0 | . | Netherlands | |
| 29 | Russian Physics Journal | 219 | 174 | 5.9 % | Russia | |
| 30 | International Journal Of Economics And Financial Issues | 215 | 0 | . | Turkey | |
| 31 | Inorganic Materials | 215 | 184 | 4.0 % | Russia | |
| 32 | Russian Metallurgy Metally | 212 | 157 | 7.8 % | Russia | |
| 33 | Quantum Electronics | 201 | 162 | 5.5 % | Russia | |
| 34 | Measurement Techniques | 197 | 130 | 11.0 % | Russia | |
| 35 | Russian Journal Of Applied Chemistry | 197 | 266 | -7.2 % | Russia | |
| 36 | Journal Of Experimental And Theoretical Physics | 196 | 200 | -0.5 % | Russia | |
| 37 | Journal Of Surface Investigation | 188 | 169 | 2.7 % | Switzerland | |
| 38 | Communications In Computer And Information Science | 185 | 0 | . | Germany | |
| 39 | Journal Of Communications Technology And Electronics | 182 | 96 | 17.3 % | Russia | |
| 40 | Mathematical Notes | 179 | 159 | 3.0 % | Russia | |
| 41 | Advances In Intelligent Systems And Computing | 178 | 0 | . | Russia | Discon-tinued |
| 42 | Mendeleev Communications | 176 | 113 | 11.7 % | Russia | |
| 43 | Journal Of High Energy Physics | 174 | 141 | 5.4 % | Russia | |
| 44 | Doklady Mathematics | 174 | 205 | -4.0 % | Russia | |
| 45 | Physics Of Atomic Nuclei | 166 | 143 | 3.8 % | Russia | |
| 46 | Russian Journal Of Physical Chemistry B | 165 | 111 | 10.4 % | Russia | |
| 47 | Journal Of Structural Chemistry | 165 | 120 | 8.3 % | Russia | |
| 48 | Physics Letters Section B Nuclear Elementary Particle High Energy Phys | 165 | 194 | -4.0 % | Netherlands | |
| 49 | Automation And Remote Control | 164 | 163 | 0.2 % | Russia | |
| 50 | Physical Review Letters | 160 | 267 | -12.0 % | USA | |